\documentclass{article}
\usepackage{spconf,amsmath,graphicx,acronym,balance}
\usepackage{url}
\usepackage[subtle]{savetrees} 
\usepackage[table,dvipsnames]{xcolor}
\usepackage{enumitem}
\setitemize{noitemsep,topsep=0pt,parsep=0pt,partopsep=0pt}

\acrodef{STFT}{short-time Fourier transform}
\acrodef{DRTF}{direct relative transfer function vector}
\acrodef{DRR}{direct-to-reverberation ratio}

\title{DBnet: DOA-driven beamforming network for \\ end-to-end farfield sound source separation}

\name{%
    Ali Aroudi$^{\dagger\star}$, Sebastian Braun$^{\dagger}$
    }
\address{%
    $^{\dagger}$ Microsoft Corporation Redmond, WA, USA \\
    $^{\star}$ Department of Medical Physics and Acoustics, University of Oldenburg, Germany \\{\tt ali.aroudi@uni-oldenburg.de, sebastian.braun@microsoft.com} 
        }    
\begin{document}
\ninept
\maketitle

%
\begin{abstract}

Many deep learning techniques are available to perform source separation and reduce background noise. However,  designing an end-to-end multi-channel source separation method using deep learning and  conventional acoustic signal processing techniques still remains challenging. In this paper we propose a direction-of-arrival-driven beamforming network (DBnet) consisting of direction-of-arrival (DOA) estimation and beamforming layers for end-to-end source separation. 
We propose to train DBnet using loss functions that are solely based on the distances between the separated speech signals and the target speech signals, without a need for the ground-truth DOAs of speakers. 
To improve the source separation performance, we also propose end-to-end extensions of DBnet which incorporate post masking networks. 
We evaluate the proposed DBnet and its extensions on a very challenging dataset, targeting realistic far-field sound source separation in reverberant and noisy environments. 
The experimental results show that the proposed extended DBnet using a convolutional-recurrent post masking network outperforms state-of-the-art source separation methods. 

\end{abstract}

%
\section{Introduction}
\label{sec:introduction}
Environmental noise, reverberation and interfering sound sources negatively affect the quality of the speech signals received at the microphones and therefore degrade the performance of many speech communication systems including automatic speech recognition systems, hearing assistive devices and mobile devices. 
In recent years several deep learning techniques have been proposed to separate out the speakers from the microphone signals and reduce background noise based on, e.g., the frequency domain transformation \cite{Kolbak_2017_IEEE_ACM_ASLP, Chen_ICASSP_2017, Kinoshita_2018_ICASSP, WHAMR_ICASSP_2020} or a learned latent domain transformation \cite{WHAMR_ICASSP_2020, TaSNet_Luo_ICASSP2018, Luo_2019_IEEE_ACM_ASLP, FurcaNet_2019}. 
The frequency-domain-based techniques perform sound source separation typically by estimating time-frequency masks corresponding to each source, while the latent-domain-based techniques aim at learning a latent space from the time-domain signals to perform source separation. 
Although most of these techniques are able to separate out speakers from single-channel microphone signals, designing a multi-channel source separation method that properly leverages all inter-channel information remains to be solved. 

When multi-channel inputs are available or a physical interpretation of a signal is possible conventional acoustic signal processing, e.g., beamforming and direction-of-arrival estimators (DOA), have analytical solutions and reasonably good performace in many cases. 
This motivates to integrate conventional acoustic signal processing techniques and deep learning techniques to profit from both worlds, as has been proposed by several works \cite{Drude_2019_JSTSP, Yoshioka_ICASSP_2019, Ochiai_2020_ICASSP}.  
However, the integration of techniques is typically performed in a modular way where each module is optimized individually, which may lead to non-optimal solution. 
More recently, a filter-and-sum network with an end-to-end training has been proposed using time-domain filtering, which does not benefit from computationally efficient  filtering in the frequency domain \cite{Luo_ASRU_2019}. 
In this paper we propose a DOA-driven beamforming network (DBnet) aiming at end-to-end sound source separation where the gradient is propagated in an end-to-end optimization way from time-domain separated speech signals to time-domain microphone signals through layers operating in the frequency and latent domains. 
As opposed to most existing techniques in literature, we evaluate the proposed  DBnet on a very challenging and realistic large-scale dataset, targeting realistic far-field sound source separation in reverberant and noisy environments.



\begin{figure}[t]
 \centering
  \centerline{\includegraphics[width=8.5cm]{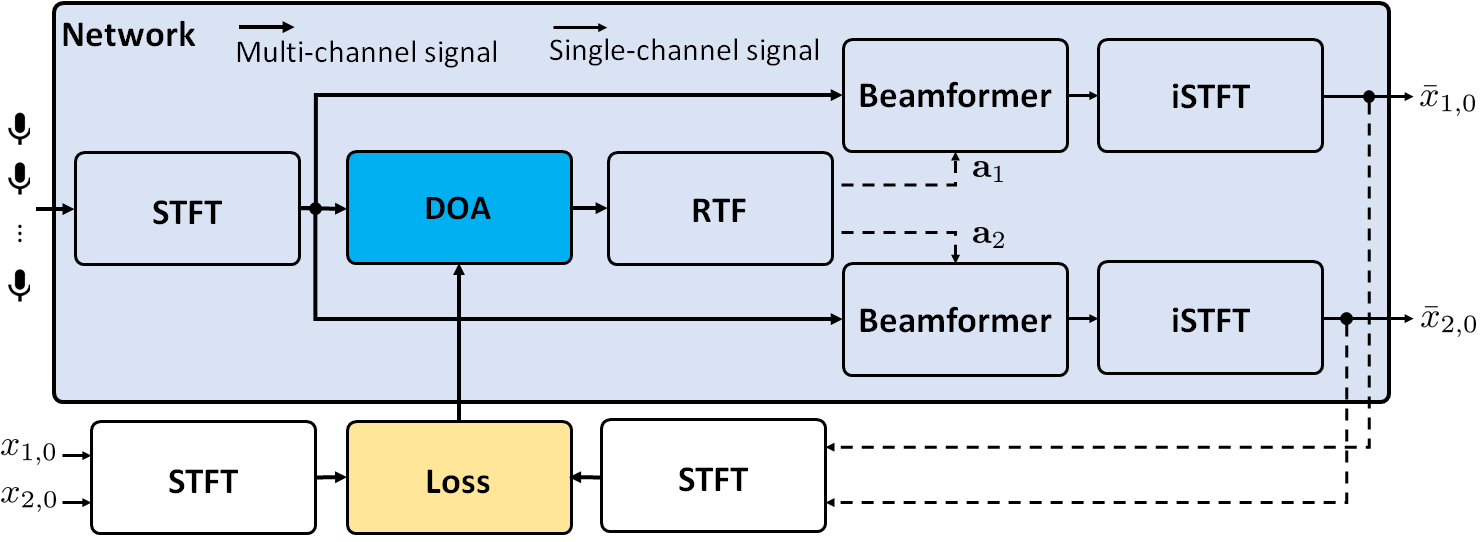}}
 \caption{\small Block diagram of the proposed DBnet structure}
\label{fig:DBnet}
\end{figure}

The proposed DBnet is depicted in Fig. \ref{fig:DBnet}. DBnet consists of layers for short-time Fourier transform (STFT), DOA estimation, beamforming and inverse STFT (iSTFT). 
We train DBnet using loss functions that are solely based on the distances between the separated speech signals and the target speech signals. In this way, the network is trained to inherently estimate the DOAs of speakers leading to the  best possible separated speech signals while the need of the ground truth DOAs for training the network is eliminated.
Although DBnet is able to separate the sources and suppress noise, the output signals may still contain residual noise. Therefore, we also propose end-to-end extensions of DBnet which incorporate a post masking network (pMnet).  For pMnet we consider a convolutional-recurrent network with an encoder-decoder  architecture which combines frequency and latent-domain based techniques. 
We train the networks using several loss functions based on complex spectra distances, magnitude distances or a combination of both distances.
In addition, aiming at developing the networks to be well-generalized for unseen acoustic conditions,  we generate a large-scale training dataset using several augmentation steps. 


We experimentally compare the proposed DBnet and DBnet extensions with the state-of-the-art source separation methods for challenging noisy and reverberant conditions. The results show that the proposed extended DBnet outperforms state-of-the-art source separation methods. 

%
\section{End-to-end sound source separation}
\label{sec:End-to-end sound source separation}

\subsection{Problem formulation}
\label{subsec: Problem formulation}
We consider an acoustic scenario comprising two competing speakers and background noise in a reverberant environment. We consider an array with $M$ microphones. The sound captured at the $m$-th microphone signal can be decomposed as
\begin{equation}
    y_{m}\left[n\right]=\overset{2}{\underset{i=1}{\sum}}x_{i,m}\left[n\right]+v_{m}\left[n\right],
\end{equation}
where $x_{i,m}\left[n\right]$ denotes the direct speech component in the $m$-th microphone signal corresponding to speaker $i$, $v_{m}\left[n\right]$ denotes the noise component representing reverberation, background noise and any remaining components and $n$ denotes the discrete time index. 

In the \ac{STFT} domain, the $M$-dimensional stacked vector of all microphone signals can be given by
\begin{equation}
    \mathbf{y}\left(k,f\right)=\overset{2}{\underset{i=1}{\sum}}\mathbf{a}_{i}\left(k,f\right)X_{i,0}\left(k,f\right)+\mathbf{v}\left(k,f\right),
\end{equation}
where $\mathbf{a}_{i}\left(k,f\right)$ denotes the \ac{DRTF} corresponding to source $i$, $X_{i,0}$ denotes the STFT coefficient of $x_{i,0}\left[n\right]$,  $\mathbf{u}\left(k,f\right)$ denotes the noise component, and $k$ and $f$ are the frame index and the frequency index. Assuming the sources are in the far field of omnidirectional microphones, the \ac{DRTF} $\mathbf{a}_{i}\left(k,f\right)$ can be given as 
\begin{equation}
    \mathbf{a}_{i}(k,f)=e^{j\kappa(k)\mathbf{R}_{\text{mic}}\mathbf{r}_{i}},
    \label{eq: RTF}
\end{equation}
where $\mathbf{R}_\text{mic}$ denote the $M\times 3$ Cartesian microphone array coordinates, $\mathbf{r}_i$ is the Cartesian $i$-th source position, normalized to unit distance, and $\kappa(k) = \frac{2\pi k f_s}{c N_\text{FFT}}$ is the wavenumber with $c$ denoting the speed of sound, $f_s$ denoting the sampling rate, and $N_\text{FFT}$ denoting the fast Fourier transform (FFT) size. 
With the far field assumption, the position of $i$-th source $\mathbf{r}_i$ in the far field in (\ref{eq: RTF}) can be assumed as distance-independent, and therefore only depends on elevation and azimuth angles,
$\theta_{\textrm{el,}i}$ and $\theta_{\textrm{az,}i}$, corresponding to the $i$-th source. 

Our goal is to separate out the direct speech signal components of the speakers from the microphone signals in the time domain using a network $h$ with linear and non-linear layers operating in the frequency and latent domains, i.e.,
\begin{equation}
    \bar{x}_{i,0}\left[n\right]=h\left(y_{m}\left[n\right]\right).
\end{equation}


\subsection{DBnet}
\label{subsec: DBnet}
The DBnet accepts microphone signals $y_{m}\left[n\right]$ as input signals and computes their corresponding STFTs using the first layer (see Fig. \ref{fig:DBnet}). The DOA estimation layer with learnable parameters then estimates the DOAs of both sources from the phase spectrograms of input signals. Based on the estimated DOAs, the \acp{DRTF} of sources are estimated using \eqref{eq: RTF} to steer two parallel beamforming layers separating out the speech sources. The output signals of the beamformers are finally transformed to the time domain using the iSTFT layer. 
We explore either convolutional-recurrent (DBnet(CR)), or fully recurrent structures (DBnet(R)) for DOA estimation, described in the following.

\begin{figure}[t]
 \centering
  \centerline{\includegraphics[width=5cm]{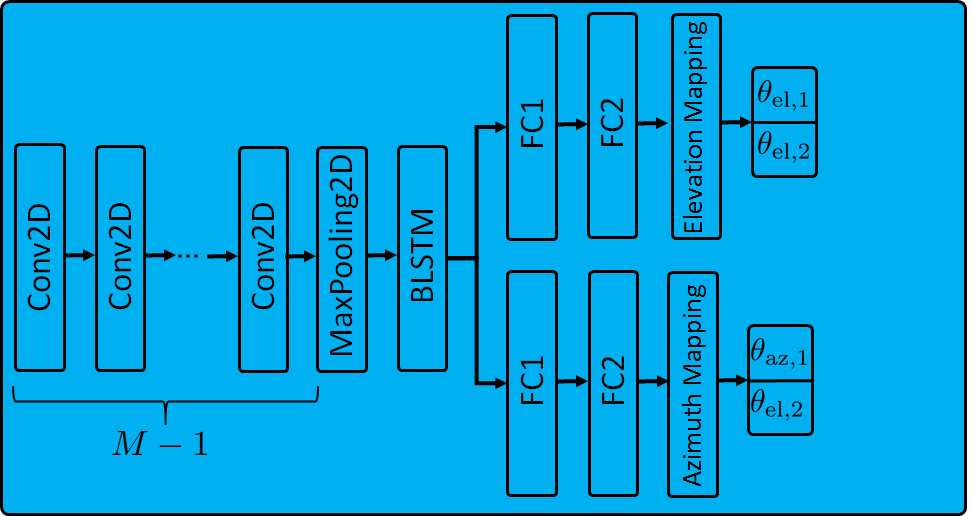}}
 \caption{\small Structure of the convolutional-recurrent DOA estimator}
\label{fig:DOA}
\end{figure}

The first DOA estimation architecture is a convolutional-recurrent structure (DBnet(CR)), to learn the relevant features from phase spectrograms and to model the sequence of the learned features  as shown in Fig.~\ref{fig:DOA}. We use $M-1$ 2D-convolutional layers with $64$ filters of size $m\times f=2\times1$ to learn the phase correlations between adjacent microphones at each frequency sub-band separately as proposed in \cite{Soumitro_IEEE_JSTSP_2019_Feb}. These learned features are then pooled using a 2D max pooling layer to increase the robustness of features against background noise. The sequence of these features is modeled using a BLSTM layer \cite{Hochreiter_1997} with $1200$ units. The outputs of the BLSTM layer then go through two parallel fully connected layers with Sigmoid functions followed by two mapping blocks to estimate the azimuth and elevation angles of the DOAs of both sources. The azimuth and elevation mapping blocks map the output of the fully connected layers to azimuth angle  $\theta_{\textrm{el,}i}\in\left[-175^{\circ},185^{\circ}\right]$ and elevation angle $\theta_{\textrm{az,}i}\in\left[-175^{\circ},185^{\circ}\right]$ ranges, respectively. 

The second DOA estimation architecture is a fully recurrent (DBnet(R)) structure, consisting only of the BLSTM, 2D max pooling, and fully connected layers of the convolutional-recurrent structure. For both DOA estimation architectures, all 2D-convolutional layers use strides of $k\times f=1\times1$ and the 2D max pooling layer uses strides of $32\times32$, downsampling across time and frequency dimensions. The BLSTM layers is an aggregation step returning only the last hidden state, resulting in time-and-frequency-invariant outputs.

For beamforming, we use the linearly-constrained-minimum-variance (LCMV) beamformer \cite{Veen1988}, minimizing spatially diffuse noise, while preserving the target source signal and spatially nulling the interfering source. The LCMV beamformer is computed from the \acp{DRTF} corresponding to the estimated source angles $\{\theta_{\textrm{el,}i}, \theta_{\textrm{az,}i}\}$ using  \eqref{eq: RTF}, and the isotropic noise field covariance matrix \cite{Borisagar_2001}.





\setlength{\textfloatsep}{6pt}

\setlength{\textfloatsep}{6pt}
\subsection{DBnet extensions with post masking}
\label{subsec: DBnet with post masking}
\begin{figure}[t]
 \centering
  \centerline{\includegraphics[width=6cm]{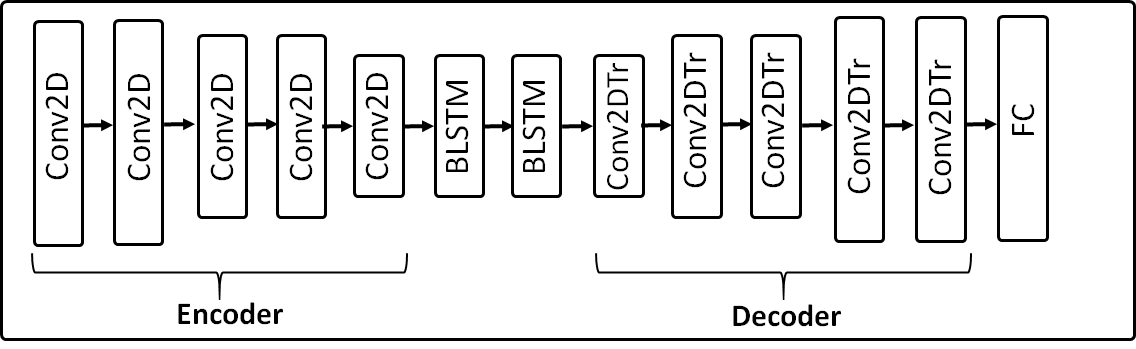}}
 \caption{\small Structure of the convolutional-recurrent pMnet}
\label{fig:pMnet}
\end{figure}


\begin{table*}[t]
\caption{\small Loss functions}
\centering
\renewcommand{\arraystretch}{.5}
\begin{tabular}{ccccc}
\hline
\rowcolor[gray]{.9}
\bfseries Loss function & Complex ($\mathcal{L}_{\mathrm{complex}}$) & & Magnitude ($\mathcal{L}_{\mathrm{magnitude}}$) &\\
\hline\hline

MSE & $\log_{10}\underset{k,f}{\sum}\left\Vert X_{i,0}\left(k,f\right)-\overline{X}_{i,0}\left(k,f\right)\right\Vert ^{2}$ & (3) & $\log_{10}\underset{k,f}{\sum}\left\Vert \left|X_{i,0}\left(k,f\right)\right|-\left|\overline{X}_{i,0}\left(k,f\right)\right|\right\Vert ^{2}$ & (4) \\

\rowcolor[gray]{.9}
cMSE & $\log_{10}\underset{k,f}{\sum}\left\Vert \left|X_{i,0}\left(k,f\right)\right|^{c}e^{j\varphi_{X}}-\left|\overline{X}_{i,0}\left(k,f\right)\right|^{c}e^{j\varphi_{\overline{X}}}\right\Vert ^{2}$ & (5) & $\log_{10}\underset{k,f}{\sum}\left\Vert \left|X_{i,0}\left(k,f\right)\right|^{c}-\left|\overline{X}_{i,0}\left(k,f\right)\right|^{c}\right\Vert ^{2}$ & (6)\\ 

MAE & $\log_{10}\underset{k,f}{\sum}\left\Vert X_{i,0}\left(k,f\right)-\overline{X}_{i,0}\left(k,f\right)\right\Vert $ & (7) & $\log_{10}\underset{k,f}{\sum}\left\Vert \left|X_{i,0}\left(k,f\right)\right|-\left|\overline{X}_{i,0}\left(k,f\right)\right|\right\Vert $ & (8) \\

\rowcolor[gray]{.9}
SDR & $-\log_{10}\frac{\underset{k,f}{\sum}\left\Vert X_{i,0}\left(k,f\right)\right\Vert ^{2}}{\underset{k,f}{\sum}\left\Vert X_{i,0}\left(k,f\right)-\overline{X}_{i,0}\left(k,f\right)\right\Vert ^{2}}$ & (9) & $ $ & \\

\hline
\end{tabular}
\label{tab: Loss functions}
\vspace{-4mm}
\end{table*}

Since the suppression capability of linear spatial filters such as the LCMV beamformers used in DBnet is limited,  
the output signals of the DBnet may still contain residual noise. Therefore, we  propose an extension of DBnet, which incorporates a post masking-based network (pMnet) to suppress the residual noise. pMnet takes the log magnitude spectrum of the beamformer output signals and generates real-valued time-frequency masks corresponding to each source. 
Motivated by the results in \cite{Kolbak_2017_IEEE_ACM_ASLP, WHAMR_ICASSP_2020} where a masking-based network with recurrent layers was used for speech separation, we consider a recurrent pMnet consisting of four BLSTM layers followed by one fully connected layer to estimate the stacked masks corresponding to the sources. As an alternative to the recurrent pMnet, we also consider a convolutional-recurrent pMnet with an encoder-decoder network architecture (see Fig. \ref{fig:pMnet}), similarly as used in \cite{Tan_ICASS_2019} but for speech enhancement. The 2D-convolutional layers encode the input magnitude spectrums into a higher-dimensional latent features and two BLSTM layers model the sequence of latent features. The outputs of the BLSTM layers are converted back to the original input dimension using mirrored transposed 2D-convolutional layers. The output of the encoder-decoder layers go then through a feed forward layer to estimate the masks of both sources. For both recurrent and convolutional-recurrent pMnets, the BLSTM layers have $1200$ units and the fully connected layers have $514$ units with Sigmoid functions. For the convolutional-recurrent pMnet,  the sequence of 2D-convolutional layers has 16, 16, 32, 32, 64 filters with kernel sizes of $k\times f=6\times6$ and strides of $1\times2$. The sequence of transposed 2D-convolutional layers correspondingly has 32, 32, 16, 16, 2 filters. All convolutional layers are followed by leaky ReLU functions.

To investigate the impact of the post masking on the source separation performance, we consider the following end-to-end DBnet extensions: 
\begin{itemize}[noitemsep,topsep=0pt,parsep=0pt,partopsep=0pt]
    \item \textbf{DBnet(R)-pMnet(R)}: recurrent DBnet followed by the recurrent Mnet
    \item \textbf{DBnet(R)-pMnet(CR)}: recurrent DBnet followed by the convolutional-recurrent pMnet
    \item \textbf{DBnet(CR)-pMnet(R)}: convolutional-recurrent DBnet followed by the recurrent pMnet
    \item \textbf{DBnet(CR)-pMnet(CR)}: convolutional-recurrent DBnet followed by the convolutional-recurrent Mnet
\end{itemize}

\subsection{Baseline method: masking-based source separation}
\label{subsec: Masking-based source separation}
As baseline methods we consider a recurrent masking-based network (Mnet(R)) and a convolutional-recurrent masking-based network (Mnet(CR)). 
The Mnet(R) has a similar structure as used for pMnet(R) with only one difference that the log magnitude spectrum of a reference microphone signal is used as input. 
The Mnet(CR) uses the same input and has a similar structure as used for pMnet(CR). The baseline methods use only the reference microphone from the array.

\section{Loss functions}
\label{sec: Loss functions}
For training the networks we consider several loss functions using either complex spectral distance or magnitude distance between the STFT of the separated speech signals $\overline{X}_{i,0}\left(k,f\right)$ and the STFT of the target speech signals $X_{i,0}\left(k,f\right)$, given in Table \ref{tab: Loss functions}. We consider four different loss functions in the following: i) the mean squared error (MSE) given by (3) and (4) \cite{Braun2020_DnnLosses}; ii) the compressed MSE (cMSE) given by (5) and (6), which is computed based on the magnitudes compressed with an exponent $c=0.3$  \cite{Lee_2018_IEEESPL, Wilson_2018_IWAENC}, in order to deal with the large dynamic ranges of audio signals;
iii) the mean absolute error (MAE) loss functions given by (7) and (8), promoting speech sparsity \cite{Braun2020_DnnLosses};  iv) 
the commonly used scale-variant signal-to-distortion ratio (SDR) (9) \cite{Roux_2019_ICASSP, Ochiai_2020_ICASSP}.

In addition, we consider loss functions which combine the complex spectral distance and the magnitude distance per row of Table \ref{tab: Loss functions}, i.e., $\mathcal{L}=\alpha\mathcal{L}_{\mathrm{complex}}+\left(1-\alpha\right)\mathcal{L}_{\mathrm{magnitude}}$ where $0\leq\alpha\leq1$ denotes the loss combining factor. 
The source-to-output mapping problem is solved by using utterance permutation invariant training (uPIT) as proposed in \cite{Kolbak_2017_IEEE_ACM_ASLP} is employed.

\section{Datasets}
\label{sec: Data augmentation}
Since the performance of neural networks are highly data dependent, we put large effort in building realistic and large enough training and test sets to draw valid conclusions. We separate training, validation and test data as much as possible using different datasets where possible. Since room impulse responses (RIRs) for our targeted microphone array are unfeasible to obtain from measurements in large quantity, we simulate RIRs for training, while we use measured RIRs for validation and testing to ensure generalization of the netrworks to real-world acoustic conditions.

\subsection{Training, validation and test data}
\label{subsection: training, validation and test data}
For training, validation and testing, we use three different speech databases, i.e.\,
540~h of speech data from audiobooks rated with high quality published in the Deep Noise Suppression Challenge \cite{data_DNS-Challenge}, 18~h from VCTK \cite{VCTK}, and 5~h from DAPS \cite{DAPS}, respectively.


We consider a 7-channel microphone array with 6 microphones on a circle of 4~cm radius and one center microphone. The center mic $m=0$ is defined as reference microphone. For training, the array geometry is simulated as free-field omnidirectional microphones, while for validation and testing we use measured RIRs using an actual device. 
For training, we simulate RIR sets of random positions in 1000 differently sized rooms using the image method \cite{Allen1979}, while for validation and testing we use measured RIRs using the actual device in 6 different rooms. The rooms were office, meeting, living rooms, and a large entrance hall with reverberation times between 0.3 to 1.3~s. In each room, several source positions were measured at challenging distances between 2 to 10~m, resulting in \acp{DRR} between -10 to 0~dB.

To generate spatially realistic noise, we use an internal database of recordings made with a third-order Ambisonics array, which were rendered to the 7-channel microphone array using a full spherical set of \acp{DRTF} from all source positions to the microphone positions. For training, we use a subset of 39~h of the noise, rendered accordingly to \acp{DRTF} simulated with the image method, while for validation and testing data we use two unseen 2.5~h noise subsets, which were rendered to measured \acp{DRTF} of the actual device in an anechoic chamber.

\subsection{Data generation}
For data generation we use the same processing pipeline for all datasets:
We create mixtures of two overlapping speech signals of 30~s length. Each source signal is generated by concatenation of recordings from the same speaker, convolved with a RIR from a randomly chosen position in the same room. The reverberant multichannel speech signals are mixed with energy ratios drawn from a normal distribution with 0~dB mean and 1~dB variance. Noise is added to the mixtures with SNRs drawn from a normal distribution with mean and variance of 8~dB and 10~dB, respectively. The microphone signals are then generated by scaling the reverberant-noisy mixtures with levels from a normal distribution with mean and variance of -28~dB and 10~dB. The target speech signals are generated using windowed versions of reverberant RIRs, enforcing a maximum reverberation time of 200~ms. These target speech signals are also scaled jointly with the microphone signals. 
Using this pipeline, we generate training, validation and test sets of 1000~h, 2.5~h and 4~h, respectively.
All datasets were generated at a sampling rate of $f_{s}=16$ kHz. 

\section{Network training and STFT setup}
\label{sec: Network training and STFT setup}
All DOA estimation and beamforming layers were implemented using a weighted overlap-add framework with an STFT frame length of 512 samples, an overlap of $50\%$ between successive frames, a Hann window and an FFT size $N_\text{FFT}=512$. 

All networks were trained using Adam optimizer \cite{Kingma2014}. 
The initial learning rate was set to an appropriate value and then decreased by a factor of 2 if the SDR validation loss does not improve for 2 consecutive epochs. 
In each epoch, 500 batches of 10 audio sequences were randomly selected from the training set. Each sequence was a random 10~s sub-portion from the 30~s signals.  
To improve the network generalization, we use gradient clipping technique with a maximum $L_{2}$ norm of 5, similarly as used in \cite{WHAMR_ICASSP_2020}.


\section{Experimental results}
\label{sec: Experimental results}
In this section, we evaluate the speech separation performance of the proposed networks in terms of the scale-invariant SDR and SIR of BSSEval \cite{BSSEVAL} and PESQ \cite{PESQ}. 
In Section \ref{subsection: Loss functions and source separation performance}, we investigate the impact of loss functions and network structures on the performance of the masking-based networks. 
In Section 5.2, we investigate
the impact of using post masking on the performance of DBnets.

\subsection{Loss functions and source separation performance}
\label{subsection: Loss functions and source separation performance}
\begin{figure}[t]
 \centering
  \centerline{\includegraphics[width=6cm]{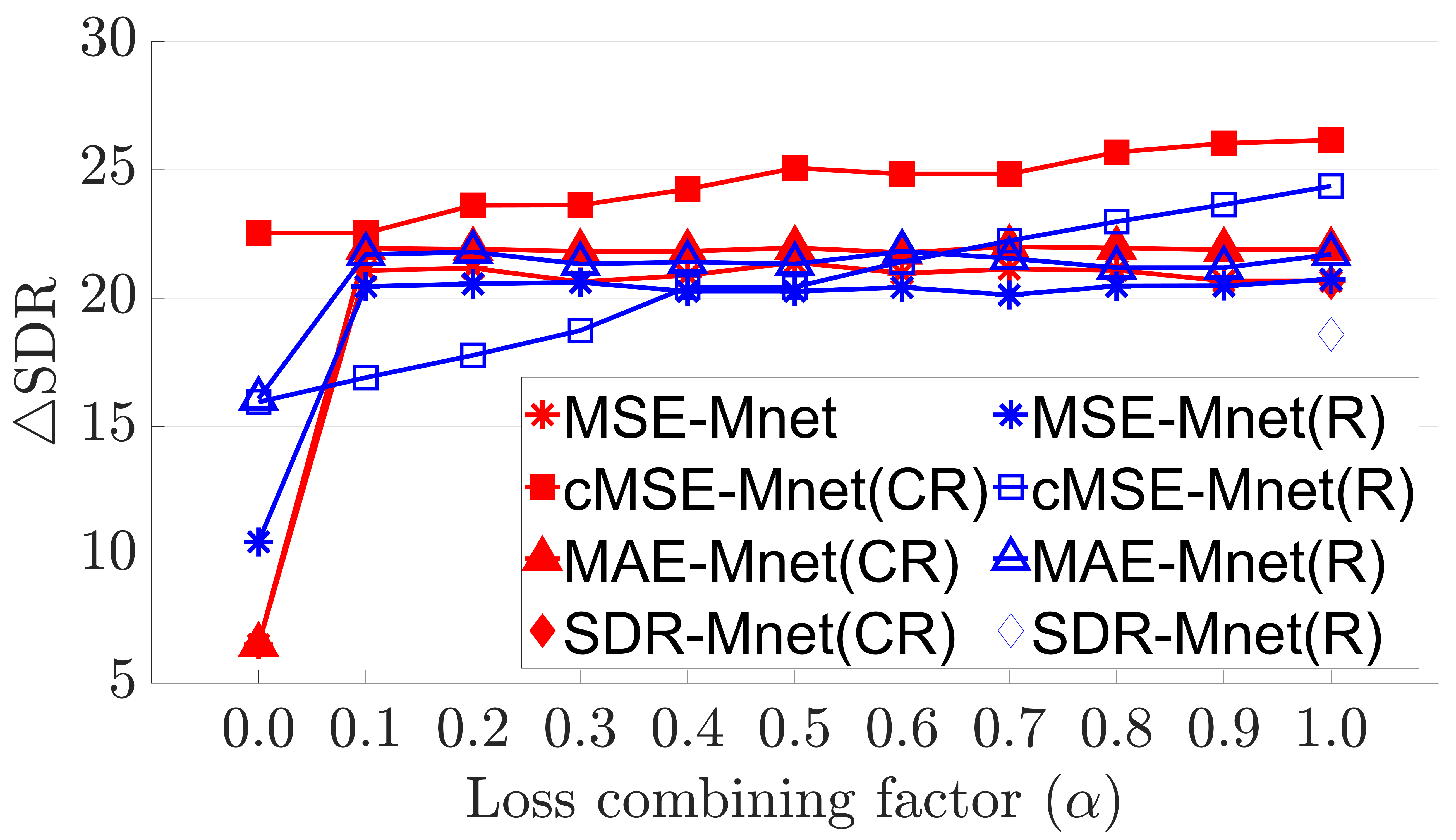}}
 \caption{\small SDR improvement for different loss functions when using Mnet(R) and Mnet(CR) }
\label{fig:SDR}
\end{figure}

Figure \ref{fig:SDR} depicts the SDR improvement of the masking-based networks either using the recurrent structure (Mnet(R)) or the convolutional-recurrent structure (Mnet(CR)) for all considered loss functions. 
We observe for all loss functions that the lowest SDR improvement is obtained when training the networks based on only the magnitude distance, i.e., $\alpha=0$, and the largest SDR improvement is obtained when training based on the complex spectral distance, i.e., $\alpha=1$. In addition, training the networks based on the cMSE loss function yields the highest SDR performance, showing the importance of dynamic range compression for network training. Therefore, from now on we focus only on the results obtained based on the cMSE loss function. In Table \ref{tab: Mnet} the SDR, the SIR and the PESQ improvement of the masking-based networks are compared. We observe that the Mnet(R) tends to degrade the SIR improvement, while the Mnet(CR) yields a significant improvement in terms of all considered performance measures, showing the effectiveness of convolutional-recurrent structure for masking based source separation. 

\subsection{DBnet source separation performance}
In Table \ref{tab: DBnet} the source separation performance of DBnets and their extensions are compared. 
We observe that the DBnets either using the recurrent structure (DBnet(R)) or the convolutional-recurrent structure (DBnet(CR)) provide an SDR and an SIR improvement, however the SDR improvement is lower than the Mnets (Table \ref{tab: Mnet}). The lower SDR improvement of DBnets can be mainly attributed to the limited suppression capability of beamformers, resulting in output signals with residual noise. 
Nevertheless, both DBnet(R) and DBnet(CR) are still able to yield a larger SIR improvement of about $0.82-1.26$ dB compared to the Mnet(R) (Table \ref{tab: Mnet}). The SIR improvement implies that DBnet(R) and DBnet(CR) are able to estimate the DOAs of the sources to appropriately steer beamformers, although both networks were trained without using the ground-truth DOAs of sources. This confirms the possibility of training beamforming networks without having ground-truth DOAs for source separation. 


When using the extensions of DBnets using post masking, it can be observed that all considered performance measures are significantly improved compared to the DBnets, showing the importance of post masking. 
When using the convolutional-recurrent DBnets followed by post masking, a considerably larger SIR improvement of $3.02$ dB for DBnet(CR)-Mnet(R) and $6.86$ dB for DBnet(CR)-Mnet(CR) is obtained compared to the Mnet(CR). 
However, when using the recurrent DBnets followed by post masking (DBnet(R)-Mnet(R) and DBnet(R)-Mnet(CR)), the improvement for all measures decreases. In general, among all considered networks the Mnet(CR) yields the highest SDR improvement but low SIR improvement and  DBnet(CR)-Mnet(CR) yields the highest SIR and PESQ improvement with a considerably high SDR improvement.

\begin{table}[t]
\caption{\small Comparison of Mnet with recurrent and convolutional-recurrent structures}
\centering
\renewcommand{\arraystretch}{.98}
\begin{tabular}{cccc}
\hline
\bfseries Method & \bfseries $\triangle\textrm{SDR}$ & \bfseries $\triangle\textrm{SIR}$ & \bfseries $\triangle\textrm{PESQ}$\\
\hline\hline

Mnet(R) & $24.36$ & $-0.21$ & $0.14$\\
Mnet(CR) & $\mathbf{26.15}$ & $2.54$ & $0.20$\\ 

\hline
\end{tabular}
\label{tab: Mnet}
\end{table}

\begin{table}[t]
\renewcommand{\arraystretch}{.98}
\caption{\small Comparison of DBnet and DBnet extensions with recurrent and convolutional-recurrent structures}
\centering
\begin{tabular}{cccc}
\hline
\bfseries Method & \bfseries $\triangle\textrm{SDR}$ & \bfseries $\triangle\textrm{SIR}$ & \bfseries $\triangle\textrm{PESQ}$\\
\hline\hline

DBnet(R) & $4.26$ & $1.25$ & $0.00$\\ 
DBnet(CR) & $4.29$ & $1.26$ & $0.00$\\ 
DBnet(CR)-pMnet(R) & $25.22$ & $3.02$ & $0.14$\\ 
DBnet(CR)-pMnet(CR) & $24.11$ & $\mathbf{6.86}$ & $\mathbf{0.21}$\\ 
DBnet(R)-pMnet(R) & $22.31$ & $1.34$ & $0.09$\\ 
DBnet(R)-pMnet(CR) & $23.53$ & $0.29$ & $0.03$\\ 

\hline
\end{tabular}
\label{tab: DBnet}
\end{table}

\section{Conclusion}
\label{sec: Conclusion}
In this paper, we proposed end-to-end source separation networks combining DOA estimation, beamforming and post masking. For DOA estimation and post masking we used recurrent and convolutional-recurrent network structures. We showed the superiority of the compressed spectral loss for source separation, and showed that this solely signal-based loss is enough to train DOA driven beamformers, without training with ground truth DOAs.  
Experiments in extremely challenging and realistic acoustic conditions with source distances up to 10~m in heavily reverberant and noisy environments showed that DBnet using the convolutional-recurrent structure for both DOA estimation and post masking is able to improve the SDR, SIR, and PESQ. 
However, further work is required to improve the separation performance even more in such adverse acoustic conditions.

\balance

\bibliographystyle{IEEEbib}
\bibliography{mainRefs_2_2}

\end{document}